# Haptics-based, higher-order Sensory Substitution designed for Object Negotiation in Blindness and Low Vision: Virtual Whiskers

Junchi Feng, Giles Hamilton-Fletcher, Todd E Hudson, Mahya Beheshti, Maurizio Porfiri, John-Ross Rizzo

*Abstract*—People with blindness and low vision (pBLV) face challenges in navigating. Mobility aids are crucial for enhancing independence and safety. This paper presents an electronic travel aid that leverages a haptic-based, higher-order sensory substitution approach called Virtual Whiskers, designed to help pBLV negotiate obstacles effectively, efficiently, and safely. Virtual Whiskers is equipped with a plurality of modular vibration units that operate independently to deliver haptic feedback to users. Virtual Whiskers features two navigation modes: open path mode and depth mode, each addressing obstacle negotiation from different perspectives. The open path mode detects and delineate a traversable area within an analyzed field of view. Then, it guides the user through to the traversable direction adaptive vibratory feedback. The depth mode assists users in negotiating obstacles by highlighting spatial areas with prominent obstacles via haptic feedback. We recruited 10 participants with blindness or low vision to participate in user testing for Virtual Whiskers. Results show that the device significantly reduces idle periods and decreases the number of cane contacts. Virtual Whiskers is a promising obstacle negotiation strategy that demonstrating great potential to assist with pBLV navigation.

*Keywords*—collision avoidance, computer vision, haptics, human–machine interfaces, low-vision aid, monocular depth estimation, open-vocabulary object detection, rehabilitation, segment anything

## I. Introduction

Visual impairment is an escalating issue, with approximately 295 million individuals currently experiencing moderate to severe visual impairment and 43 million living with blindness [1]. Such impairments severely restrict mobility, leading to unemployment [2], increased dependency [3], and reduced quality of life [4], negatively impacting psychosocial well-being [5].

Unemployment perhaps poses the greatest challenge for people with blindness or low vision (pBLV), with studies indicating that up to 81% may face joblessness [6]. A major barrier to employment for pBLV is the difficulty in navigating not only to and from work but also within workplaces. This is compounded by difficulties in other public spaces such as hospitals and government facilities, perpetuating health inequities. Enhancing mobility for pBLV is crucial for improving quality of life and alleviating economic strain.

For approximately a century, the white cane has been the dominant primary mobility aid, enhancing independence and safety for pBLV [7]. However, it has limitations. Canes require physical effort, occupy one hand, need direct contact with the environment to provide perceptual gain, and can only detect obstacles up to the length of the cane itself, and its inefficiency in complex environments [8].

Electronic travel aids (ETAs) promise to address these limitations by converting visual and sensorial information into audible and tactile feedback [9]. However, existing ETA systems have not fully addressed the broad challenges of mobility [9]. Common issues include earphones that block essential sounds needed for safe navigation [10], hand-held devices that hinder fall protection [11], large, conspicuous systems that may cause discomfort, and approaches that largely use simple, lower-order control systems, leaving them on or off and nothing in between [12].

A discreet, hands-free, vibrotactile, wearable ETA offers a potential solution. An innovative example, called Impaired Smart Service System for Spatial Intelligence and Navigation (VIS$^4$ION), involves a haptic-based, higher-order sensor substitution approach provides tailored vibratory feedback through a custom waist strap [13]. VIS$^4$ION is a personal mobility solution that serves as a customizable, human-in-the-loop, sensing-to-feedback platform to deliver functional assistance in real-time [14, 15, 16, 17, 18]. In VIS$^4$ION system, the ETA is integrated with a backpack that houses power supplies and a compact computing device. An RGB camera, affixed to the shoulder strap of the backpack, captures real-time images. Images are processed by an object detection model, enabling the identification of obstacles in the wearer's path. The locations of obstacles are conveyed to the user via the vibrotactile belt, in which each vibration unit correlates to a specific area within the camera's field of view.

Despite the promise of this approach, significant challenges remain. A primary limitation is that the effectiveness of the object detection is contingent upon the range of object classes included in the model's training dataset. Objects not represented in the dataset remain undetectable, potentially compromising user safety. Given the impracticality of encompassing all conceivable object classes within the training dataset, substantial advancements are crucial to enhance the reliability and applicability of this technology.

In this paper, we introduce an innovative obstacle detection

[1]Department of Biomedical Engineering, Tandon School of Engineering, New York University, NY 11201, USA
[2]Center for Urban Science and Progress, Tandon School of Engineering, New York University, New York, NY 11201, USA
[3]Department of Mechanical and Aerospace Engineering, Tandon School of Engineering, New York University, New York, NY 11201, USA
[4]Department of Rehabilitation Medicine, NYU Langone Health, New York, NY 10016, USA

approach for haptic-based, higher-order sensory substitution. Acknowledging the inevitable presence of unknown categories in the real world, we propose two methods to generalize obstacle identification across all classes. The first method focuses on detecting a single type of object: the ground, which is universally present and typically represents an obstacle-free zone. This approach enables the differentiation between traversable and obstructed spaces, facilitating generalization across environments with diverse obstacles. To implement the concept, we developed the open path mode, which integrates an object detection model specifically tuned to identify the ground and a segmentation model to delineate traversable spaces.

The second approach, depth mode, alleviates the challenge of unrecognized obstacle types by employing a depth estimation model. This model calculates the relative distances of surrounding objects, ensuring that users are alerted to proximate objects regardless of their class/type. Together, these modes significantly enhance the adaptability of our system, promising comprehensive more navigational aids across all spaces and obstacle conditions.

The primary objective of this study was to develop a haptic-based, higher-order sensory substitution system to enhances the navigational capabilities of pBLV. The processed outputs from the two operational modes are communicated to the user through the haptic device. Our system, named Virtual Whiskers, enhances users' spatial awareness by inte- grating cutting-edge computer vision models, including zero- shot, text-conditioned object detection, and depth estimation technologies. Our approach aims to bridge the gap between traditional mobility aids and modern technological solutions, providing a robust, user-friendly system that enhances mobility and independence for pBLV. We validated our solution through user studies involving participants who were profoundly visually impaired in a structured obstacle course.

## II. METHODS

In this section, we first provide a brief overview of the VIS$^4$ION, outlining its core components and functionalities. Following this, we detail the specific hardware configurations and software setups employed in constructing Virtual Whiskers. Additionally, we describe the experimental designs utilized to evaluate the effectiveness of this innovative assistive technology.

### A. VIS$^4$ION system

As depicted in Figure 1, VIS$^4$ION is structured around four principal components. The first is a discreet wearable backpack equipped with various sensors for distance measurement and image ranging, which gather essential data about nearby obstacles and the surrounding environment. The second component is an embedded system housed within the backpack, which provides the necessary computing power and communication capabilities to process the collected data. The third component, a haptic interface, consists of a vibrotactile belt worn around the waist. This belt communicates spatial information derived from the sensors to the user in real time, allowing for immediate and intuitive feedback about the environment. The fourth and final component is a headset equipped with binaural bone conduction speakers and a microphone. This setup enables oral communication without interfering with the user's ability to hear ambient sounds, which is crucial for safety and spatial orientation. VIS$^4$ION employs two main methods for alerting users to environmental features of interest. The first method involves audible messages delivered through the bone conduction headset. This technology allows the transmission of sound directly through the bones of the skull, ensuring that the user can receive important navigational cues without compromising natural hearing. The second method utilizes the vibrotactile feedback from the waist belt. The belt segments the mapped scene into a simplified, pixelated grid, which is then communicated to the user. This method provides a tactile representation of the surroundings, helping visually impaired users navigate more confidently and safely.

Together, these innovative components and functionalities make VIS$^4$ION a cutting-edge solution in the field of assistive technologies, significantly enhancing the independence and mobility of people with visual impairments. This system not only improves their ability to navigate complex environments but also serves as a foundation for future advancements in similar technologies.

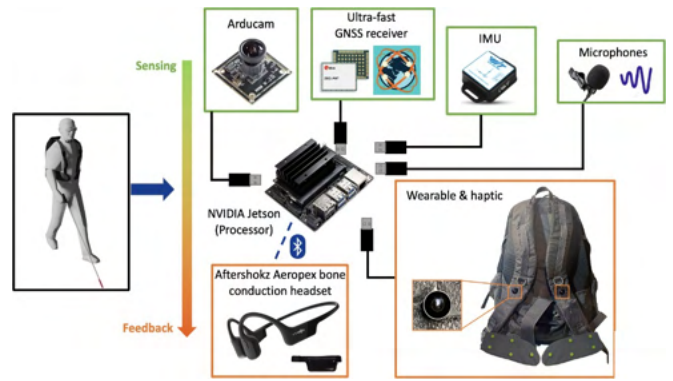

Fig. 1. The graph represents the VIS$^4$ION system and its main components. It includes several sensors such as a camera, GNSS receiver, IMU, and microphone. The processing unit is the Nvidia Jetson. The feedback components consist of a haptic feedback belt and a bone-conduction headset.

### B. Hardware Configurations

*1) VIS$^4$ION Upgrades:* Several upgrades have been performed on VIS$^4$ION in order to make it more suitable for Virtual Whiskers. The first is the embedded microcomputer, which has been upgraded to the Jetson Orin NX 16 GB to boost computational power and enhance processing capabilities. The shoulder mounted camera is Arducam 1080P Low Light Ultra Wide Angle USB Camera. Furthermore, the haptic interface via the waist strap has been re-engineered to a wireless, modular design to improve flexibility and user comfort. Detailed information about this modular design is provided in the following subsection.

*2) Modular Unit:* In the previous iteration of the VIS$^4$ION platform, the waist strap consisted of 10 vibration motors arranged in two rows of five units each. The new design maintains this 2x5 haptic feedback configuration but transitions to a modular unit approach.

*a) Components:* Each modular unit now contains two vibration motors arranged in a two-row, one-column layout. It includes 3D-printed cases, two eccentric rotating mass (ERM) vibration motors, an ESP32-based printed circuit board (PCB), and a battery. The 3D-printed components consist of a core structure fabricated from thermoplastic polyurethane (TPU), which provides structural support for the entire modular unit. The TPU 3D-printed piece can bend to some degree to better fit the user's body. The core structure features two mounting clips, as illustrated in Figure 2b. These clips enable the modular unit to be securely attached to a Nylon webbing strap. Additional components include two motor housings and two motor cases made from polylactic acid (PLA), designed to secure the motors in place and optimize contact with the user's skin through the inclusion of springs in the motor cases. The final component is a circuit cover, also printed in PLA, to protect the PCB and battery. The haptic feedback is provided by a pair of ERM vibration motors [19], positioned in a dual-row, single-column configuration. These motors are controlled by an ESP32-based PCB [20], which facilitates wireless communication with the Jetson board. The PCB features integrated Wi-Fi module for connectivity, motor drivers for operational control, voltage regulators for power management, and LED indicators for status reporting. This modular unit is powered by a 3.7 V and 1200 mAh Lithium-Ion Polymer battery [21]. Figure 2 presents a schematic illustration of the modular unit. The cost of each modular unit is around $35. To maintain the 2x5 configuration, five modular units are placed together to form a haptic feedback strap.

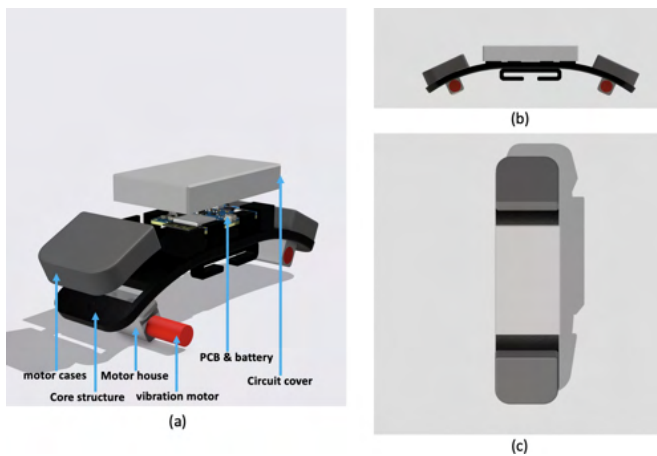

Fig. 2. 3D rendering graph of the modular unit. (a) an exploded view diagram. (b) a side view of the modular unit. (c) a top view of the modular unit.

*b) Wireless communication:* The modular units communicate with the Jetson board via a wireless network, where the Jetson board acts as a wireless access point. Each modular unit joins the network as a client, with the Jetson board serving as the host. Each modular unit has a predefined unique client ID, with the last digit of the ID serving as the position indicator. This indicator increases sequentially from left to right, such that the ID for the leftmost modular unit is client1, and the ID for the rightmost modular unit is client5. Data transmission within this network utilizes the Message Queuing Telemetry Transport (MQTT) protocol, a lightweight, publish-subscribe network protocol designed for the efficient exchange of messages between devices [22]. The software we implemented, described in a later section, publishes a motor-control signal to this network every 300 milliseconds. Each client listens for this signal and decodes it to determine whether its motors should vibrate. The motor-control signal is formatted as a list of 10 integers. Each number in the list corresponds to one vibration motor for the modular unit. The first number in the list corresponds to the top row vibration motor for client1, and the second number corresponds to the bottom row vibration motor for client1. The numerical value of the integers indicates the vibration intensity.

*c) Vibration specifications:* Each modular unit contains two vibration motors that can operate independently. Each vibration motor is programmed to have three intensity levels: high, medium, and low, which correspond to numerical values 3, 2, and 1 in the motor-control signal, respectively. The frequencies for high, medium, and low intensities are approximately 250 Hz, 150 Hz, and 80 Hz, respectively, which fall within the 40–400 Hz range in which human skin can perceive vibration [23, 24]. Every time the modular unit receives a motor triggering signal, it vibrates continuously for 100 milliseconds, followed by a 200-millisecond silent period. This setup ensures that users can discriminate vibration signals without habituation. The modular units operate independently from the Jetson board. The software running on the Jetson board takes about 150 milliseconds to process an image, which will be discussed later. During the vibration or silent period, the software continuously processes frames. Thus, the time delay between obstacle or free space detection and vibration output is up to 150 milliseconds.

*d) Spacing:* Five modular units are attached to the same nylon webbing strap with side release buckles. The nylon webbing strap is 1.5 inches wide and 55 inches long. The length is adjustable to fit the user's waist circumference. The strap should be as close to the waist circumference as possible without causing discomfort, ensuring that the modular units make contact with the user's skin. The modular units are spaced around 2 to 3 inches apart horizontally, depending on the user's body size. The central modular unit, which is the third from the left, should align with the user's center. The leftmost modular unit should align vertically with the user's left ear, and the rightmost modular unit with the right ear. The second modular unit from the left and the second from the right should be spaced evenly between the central modular unit and the leftmost and rightmost modular units, respectively. Figure 3 shows an actual haptic feedback belt.

*C. Operational Modes*

Computer vision has proven to be an effective solution with numerous successful applications in the field of assistive technology [25, 26, 27]. Therefore, for Virtual Whiskers, we have developed two distinct operational modes that leverage advanced computer vision algorithms to analyze camera footage in real-time.

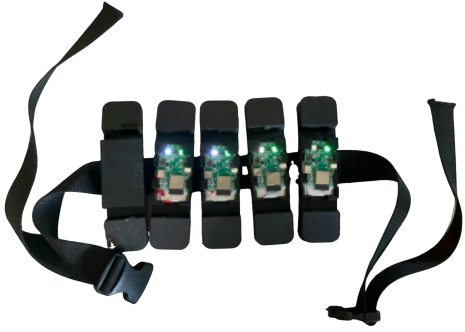

Fig. 3. A picture of the haptic feedback belt formed by five modular units attached to a nylon webbing strap with side release buckles. The leftmost modular unit is the final version, while the remaining four modular units have their circuit covers removed for photographing purposes to demonstrate the PCB. On each PCB, the left LED light indicates the battery level, and the right LED light shows the wireless connection status.

*1) Open path mode:* The open path mode is devised to discern the most spacious area in the user's immediate vicinity and direct the user towards that direction for obstacle avoidance. Open path mode involves two-staged operations. The first stage of open path mode uses an advanced object detection model called NanoOWL to identify floor objects in images. NanoOWL is based on Google's OWL-ViT model, a leading technology in object detection that can understand and process images using just text descriptions [28]. This model is built on a structure known as the Vision Transformer, which, along with specialized components for detecting and understanding objects, allows it to locate items in images using text queries alone. Nvidia has adapted NanoOWL for the NVIDIA Jetson Orin platform, enhancing its performance for real-time applications on portable devices [29]. In this setup, we configure the text prompt as "the floor" and input a video frame into the NanoOWL model. The system selects the single bounding box with the highest confidence score as the output. We establish a minimum confidence threshold of 0.02; scores below this threshold lead us to conclude that no floor is detected.

A bounding box is defined by its top-left and bottom-right coordinates. In our system, the bottom-right coordinate of the output bounding box is adjusted to match the bottom-right corner of the video frame, ensuring that the bounding box encompasses the area directly in front of the user, while the top-left corner remains unchanged. To minimize temporal inconsistencies that may arise from motion blur, variations in lighting, and occlusions, the coordinates of the bounding boxes are averaged over two consecutive frames. The resulting averaged bounding box is then forwarded to the NanoSAM model for further analysis.

NanoSAM, a specialized variant of the Segment Anything Model (SAM), marks a significant stride in image segmentation technology, pioneered by Meta [30]. Image segmentation determines which pixels in an image correspond to specific objects, a fundamental task in computer vision with applications that span from scientific imaging to photo editing. SAM's design enables it to generalize object recognition effectively, allowing it to generate segmentation masks for a diverse array of objects across different images or videos, even those outside its training scope. This capability allows the model to respond flexibly to various segmentation prompts, ranging from simple points to complex bounding boxes. NanoSAM is tailored for real-time operations on NVIDIA Jetson Orin platforms [31].

In practical application, NanoSAM's bounding box mode takes bounding box prompts to segment the most significant and cohesive object within the provided area. Specifically, when tasked with a bounding box surrounding the floor, NanoSAM adeptly segments the floor components, effectively ignoring any obstacles that might interfere, such as pillars. Figure 4 illustrates this workflow, showcasing how, even with a central pillar within the image, the segmentation of floor areas remains unaffected, demonstrating the robustness and practical utility of NanoSAM in real-world scenarios.

After segmentation, mapping the results onto the haptic feedback belt is a critical step. The belt is configured as a 2x5 grid based on the hardware setup. To facilitate calculations in subsequent steps, a margin area is added on the left, top, and right sides of this grid, expanding it to a 3x7 grid. This expanded grid aligns with the segmentation mask; the height of the grid matches the height of the segmentation mask, and the grid's width corresponds with the mask's width. This alignment enables successful mapping of traversable areas onto the haptic feedback belt. The margin area plays a crucial role in this process by reducing the representation area of the original 2x5 grid, which enhances the precision of mapping between the real-world conditions and the belt's feedback. The introduction of margin areas is particularly important as it helps adjust the mapping scores to account for edges, ensuring that users are guided away from potential hazards and obstacles, thus prioritizing safety in navigation. Figure 5 graphically illustrates this procedure.

To determine which modular unit should vibrate, the algorithm employs a set of rules. Initially, the algorithm calculates the percentage of traversable space within each cell in the 3x7 grid. This grid is shown in in Figure 5c. For analytical purposes, this proportion is quantified as a score for each cell, with values ranging from 0.00 to 1.00. A score of
1.00 indicates that the cell is completely clear and thus fully traversable, whereas a score of 0.00 denotes that the cell is entirely occupied by obstacles.

The algorithm should select the direction that is most spacious. Only considering the score of a single rectangle is not enough to determine the most spacious space. For example, the bottom left and the bottom right cells in the 2x5 grid in Figure 5c both have a score 1.00, but obvious the bottom left cell is in the most spacious area. To determine such an area, the algorithm also considers the cell's neighbors. We define an adjusted score that takes a weighted sum of the cells and its neighbors,

$$\text{Adjusted\_score} = 0.4 \times C + 0.2 \times T + 0.1 \times (L + R + TR + TL)$$

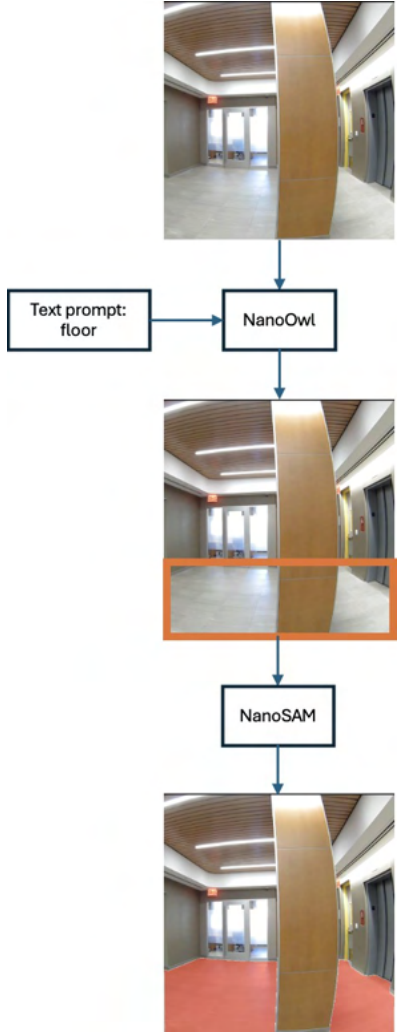

Fig. 4. Workflow for traversable space segmentation. From top to bottom: an RGB image of an indoor space with a pillar is provided to the NanoOWL model. The text prompt is "the floor." The output is a bounding box for the floor area, marked as an orange bounding box in the middle image. This image and its bounding box coordinates are then provided to NanoSAM. NanoSAM segments out the floor components in the image, as shown in the bottom image.

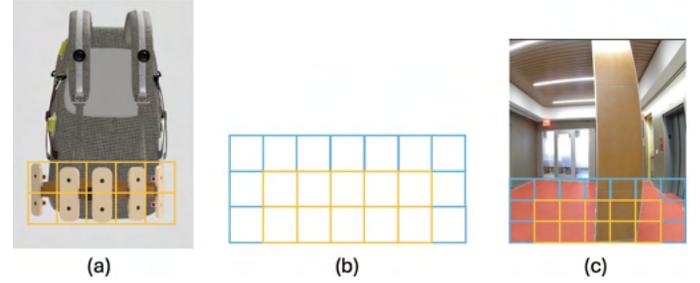

Fig. 5. The procedure for mapping the spacious area segmentation onto the haptic feedback belt. (a) The 2x5 grid, represented in yellow, corresponds to the hardware layout of the haptic feedback belt. (b) A margin area, marked in blue, is added to the left, top, and right sides of the 2x5 grid, resulting in a 3x7 grid. (c) This 3x7 grid is then overlaid onto the floor segmentation mask, ensuring that the height and width of the grid match the height and width of the floor segmentation mask.

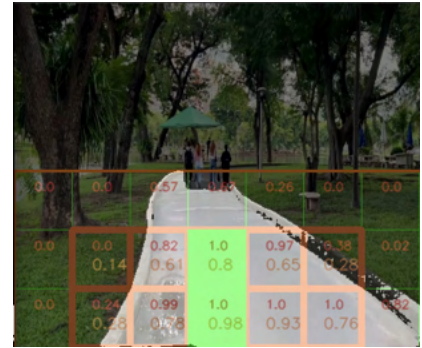

Fig. 6. An example of the final output of the open path mode of Virtual Whiskers. Here, "the floor" represents the path, which has been successfully segmented out, and the segmentation mask is visualized as white pixels. The 3x7 grid is aligned with the size of the segmentation mask. In each cell of the grid, the smaller font number at the top indicates the raw score of that cell. The larger font number in each cell of the 2x5 grid represents the adjusted score for that cell. In this example, the center column is selected, and it is marked in green.

where C = cell itself, (i,j); T = top neighbor, (i,j+1); L = left neighbor, (i-1,j); R = right neighbor, (i+1,j); TR = top right neighbor, (i+1,j+1); TL = top left neighbor, (i-1,j+1).

Note that this adjusted score is calculated solely for each cell within the 2x5 grid. Specifically, the score for the central column (third from the left) is increased by 5% to promote a central tendency in the navigation system. In scenarios where the input image is entirely traversable—where all cells uniformly achieve an adjusted score of 1.00—the central modular unit of the haptic feedback belt vibrates to signal that it is safe to proceed forward, corresponding to the central column of the grid. Additionally, any cell with an adjusted score exceeding 0.95 is similarly increased by 5% to prioritize areas that are very traversable.

Ultimately, the algorithm scans every column in the 2x5 grid and selects the single column with the highest sum of adjusted scores of the cells within it. This column represents the direction that is spacious enough to proceed. If the sum of adjusted scores for the selected column is below 0.8, no signal is generated, indicating that there is no sufficiently spacious area in the environment. If the sum exceeds 0.8 and the top cell in this column has an adjusted score of 0.9, a high-intensity vibration signal is sent to both vibration motors corresponding to the top and bottom cells. If the sum is greater than 0.8 but the top cell's adjusted score is below 0.9, only the bottom vibration motor of the corresponding modular unit receives a high-intensity vibration signal. These thresholds were empirically determined to optimize performance. Setting them below 1.00 allows the system to consider an area as traversable without requiring perfect segmentation. This is important because objects like dirt, stains, or small debris, which are common on floors but do not hinder navigation, can create gaps in the segmentation mask. The sub-1.00 thresholds ensure the system can effectively handle such scenarios. When both motors vibrate, the user feels a strong vibration, indicating that the direction is spacious enough. If only one motor vibrates, the user feels a lighter vibration, indicating that the direction is still spacious but contains some obstacles. Figure 6 is an example that demonstrates the mechanism.

*2) Depth mode:* The depth mode is designed to distinguish the presence of obstacles in the user's immediate vicinity by image depth estimation. This mode uses obstacle proximity information to direct the user towards a direction that is most obstacle free. The fundamental supposition is that it is not feasible to detect all kinds of obstacles in the real world, as there will invariably be certain categories of obstacles unknown to the object detection model. However, for obstacle negotiation, knowing the categories of obstacle is redundant. The position and distance of the obstacle is enough to avoid it. Therefore, depth estimation, the task of measuring the distance of each pixel relative to the camera, is sufficient for obstacle negotiation.

The depth mode takes a single RGB image frame as input, as shown in Figure 7 (a). This image is fed into a monocular depth estimation model, MiDas, which provides accurate relative distance measurements for all pixels in the scene [32]. The output of MiDas is a depth map of this image, as shown in Figure 7 (b). The haptic feedback belt consists of 5 modular units, each containing 2 vibration motors, forming a 2x5 grid, as depicted in Figure 7 (a). This grid is then overlaid onto the depth map, dividing the depth map into ten rectangles, as shown in Figure 7 (c). Each rectangle corresponds to one vibration motor on the belt, thereby completing the mapping between the depth map and the belt.

To determine which modular unit should vibrate, the system employs a set of rules. It is important to note that the Midas model provides only relative depth information. The system rescales the depth values to a range from 0 to 1, where 0 indicates the farthest distance and 1 the closest. The system defines relative depths greater than 0.80 as close, between 0.65 and 0.80 as medium, and between 0.50 and 0.65 as far. Values below 0.50 are ignored.

The algorithm processes each cell of the overlaid grid on the depth map by calculating the percentage of pixels classified as close, medium, or far, based on their depth relative to the user. For context, close generally refers to objects within 1 meter, medium to objects between 1 and 2 meters, and far to those beyond 2 meters. These distances were chosen based on typical user interactions within indoor environments, where immediate awareness of nearby obstacles is crucial. However, since the distance calculated by the model is relative, the actual distance represented by each classification may vary.

If more than 50% of the pixels in a cell are classified as far, the corresponding vibration motor is activated with a low-intensity signal, indicating that the area is relatively clear. When over 40% of the pixels are categorized as medium, the system adjusts to a medium-intensity vibration to signal potential obstacles at a moderate distance. If more than 30% of the pixels are classified as close, a high-intensity vibration is triggered, alerting the user to nearby obstacles. If none of these thresholds are met, the vibration motor remains inactive.

The thresholds were carefully selected through empirical testing. We found that lowering these thresholds resulted in excessive noise, reducing the system's effectiveness by overwhelming the user with unnecessary alerts. For example, when the threshold for the far category was set below 50%, the system frequently triggered vibrations in response to the floor directly in front of the user. Although these areas were indeed far from the user, they posed no immediate danger and could be safely ignored. The excessive sensitivity caused constant and distracting vibrations, making it difficult for the user to discern truly significant obstacles. This approach aims to balance sensitivity to nearby objects with a reduction in unnecessary noise, thereby enhancing overall user safety and system effectiveness.

Depth image analysis is considered complete only after commands for all rectangles have been determined. While the analysis time for each image varies slightly, all images are processed within 150 milliseconds on our device. If depth image analysis concludes in less than 150 milliseconds, the algorithm remains idle until the 150-millisecond mark is reached. Subsequently, motor control commands are sent to all actuators in the network.

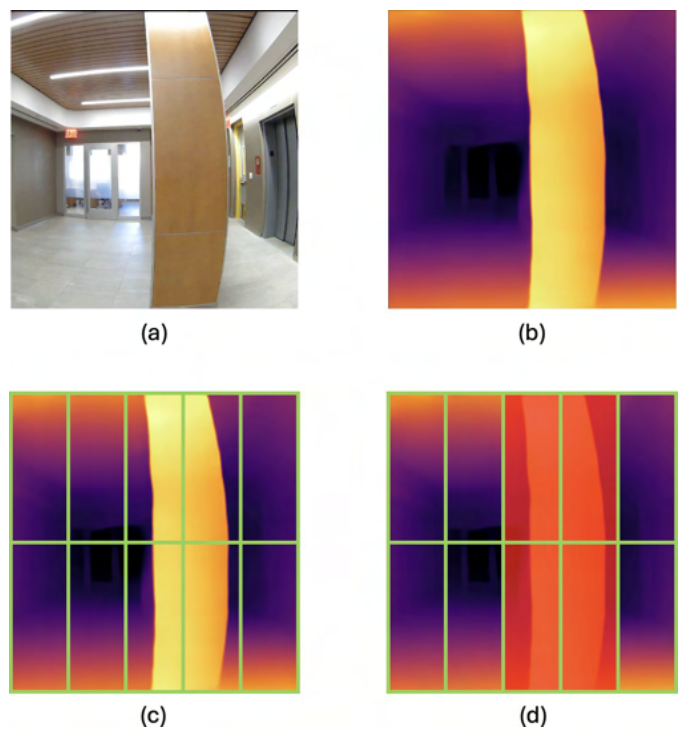

Fig. 7. Depth mode of Virtual Whiskers. (a) A RGB image frame captured from the camera of VIS$^4$ION. (b) Depth map of the corresponding RGB image generated using the Midas model. (c) The depth map is divided into a 2x5 grid, creating a one-to-one correspondence between each grid cell and a vibration motor on the belt. (d) Four cells are highlighted in red, indicating that four corresponding vibration motors are activated to signal the presence of obstacles in the user's surroundings.

## III. USER TESTING METHODOLOGY

Experiments were designed to compare user performance while navigating with a white cane alone versus navigating with both a white cane and Virtual Whiskers. Virtual Whiskers is intended as a supplement to the white cane, addressing its limitations, and is not meant to be operated independently.

## A. Participants

Participants were recruited to evaluate the effectiveness of the proposed system. The inclusion criteria encompass individuals with permanent visual impairment of varying levels and etiologies. The exclusion criteria are as follows: significant cognitive dysfunction (a score of less than 24 on Folstein's Mini Mental Status Examination), prior neurological illnesses, complex medical conditions, substantial mobility restrictions, use of walkers or wheelchairs, and pregnancy. This study received approval from the Institutional Review Board (IRB) of New York University Langone Medical Center, under the study number s17-00317.

Before participating, individuals received detailed information about the study through comprehensive discussions that outlined its objectives and procedures. After gaining a thorough understanding, participants were able to give their informed consent. Subsequently, a detailed tutorial on the two modes described earlier was provided. All participants were asked to wear occluder glasses [33], which temporarily obstruct their vision to provide a baseline of zero vision for all participants.

## B. Experimental Setup

We established an experimental field with dimensions of 70 inches by 105 inches using wall dividers. Each wall divider measures 105 inches in length and 72 inches in height. Two wall dividers were placed 70 inches apart and in parallel to each other, thereby forming the experimental field.

Safe obstacles were fabricated using swimming pool noodles and plastic vases. The pool noodles have a diameter of 6 inches and a length of 60 inches. The vase has a diameter of 8 inches and a height of 5 inches. The pool noodles are vertically inserted into the vase. The pool noodles are fixed to the vase by means of tapes. Aluminum foil was used to fill the gap between the pool noodle and the vase.

Four pool noodles and vases were placed together in this experiment to form a row of obstacles. Figure 8 is a photograph of the pool noodles in the experimental field. The principal advantage of employing pool noodles is their safety; the soft and inflatable characteristics minimize the risk of injury upon collision.

We arranged these obstacles to form an obstacle course with varying degrees of difficulty. Each difficulty level was constructed with additional rows of obstacles: one row for the easy task, two rows for the medium task, and three rows for the hard task. Figure 9 (a-c) illustrate the easy tasks, (d-f) illustrate the medium tasks, and (g-i) illustrate the hard tasks. The rows of obstacles could be either horizontal or diagonal. A horizontal row, as shown in Figure 9 a, consists of pool noodles aligned parallel to the start line of the experimental field. In contrast, the diagonal rows, as shown in Figure 9 b, consist of pool noodles arranged at a 45-degree angle to the start line. Diagonal rows are more challenging to detect because the obstacles appear narrower to the ETA.

To minimize learning effect from repeated measurements, the positions of the obstacles differ for each task. A combination of one easy, one medium, and one hard task is referred to as a block of tasks. Participants completed three blocks of tasks under three conditions respectively: 1) Virtual Whiskers open path mode with a white cane, 2) Virtual Whiskers depth mode with a white cane, and 3) white cane only. Virtual Whiskers in both modes served as the intervention or treatment, and the white cane alone constituted the control condition. The sequence in which participants utilized the devices was a cross-over design, to reduce variability among participants. Participants were encouraged to walk at a comfortable pace and were provided with a 15-30-minute tutorial on both modes of Virtual Whiskers. If participants self-reported a full understanding of the system, the tutorial could be concluded prematurely.

For each task, participants commence from the starting point, indicated by the red points in Figure 9, and proceeded to the destination, marked by the blue points in the same figure. Participants were required to walk from the beginning to the destination and then return to the starting point, making an effort to avoid obstacles along the way.

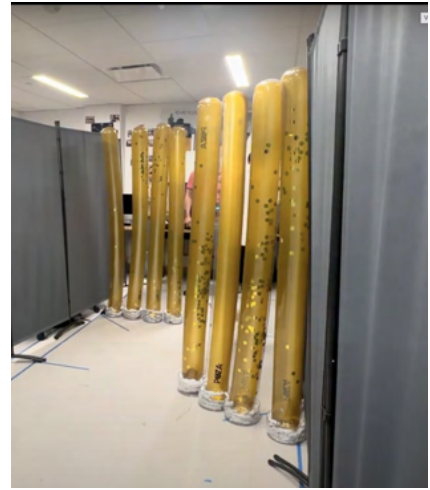

Fig. 8. A photograph of the experimental field with obstacles. The wall dividers are grey and positioned on the left and right sides of the experimental field. The obstacle consists of a yellow pool noodle placed vertically in a transparent vase. Aluminum foil, shown in silver, is utilized as filling material to fill the gap between the pool noodle and the vase.

## C. Data collection

Participants are provided with the VIS$^4$ION platform backpack and a haptic feedback belt. An iPhone mounted on the shoulder strap recorded the user's trajectory. Research in [34] provided an ARkit-based trajectory estimation app with excellent accuracy, a drift error of about 2 cm per second. Our experiment utilizes this app for trajectory recordings.

Participants used their own white canes during the experiments. Figure 10 illustrates this setup. If a participant did not bring their own white cane, an appropriately fitting white cane was provided. Two experimenters filmed the participants during the experiment for reference purposes.

We collected several data points during the experiment:
1) Total task completion time: This measures the time from when the user starts moving at the starting point until their body or white cane touched the destination line, including the time taken to walk back. Virtual Whiskers translates visual information to users via haptic feedback. We hypothesized that the use of Virtual Whiskers would decrease the task completion time compared to using a white cane only.

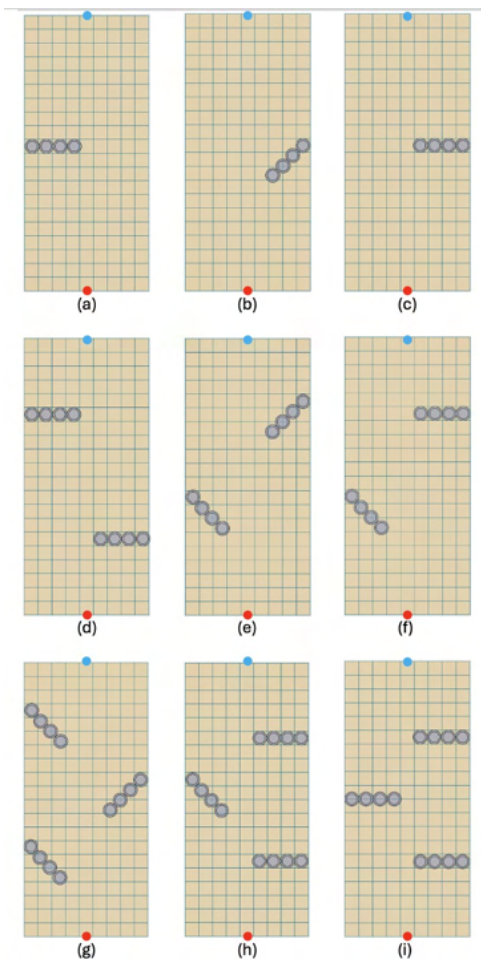

Fig. 9. Obstacle course layout. (a), (b) and (c): three simple tasks featuring a single horizontal or diagonal row of obstacles. (d), (e) and (f): three medium difficulty tasks featuring two horizontal or diagonal rows of obstacles. (g), (h) and (i): three hard difficulty tasks featuring three horizontal or diagonal rows of obstacles. The red points represent the start point for each task and the blue points represent the end point for each task.

2) Percent of hesitation time: Hesitation time is defined as the duration a participant stops moving due to obstacles, during which they swing their white cane and search for a new path. The percent of hesitation time is the ratio of hesitation time to task completion time. Our hypothesis was that the use of Virtual Whiskers would reduce the percent of hesitation time compared to using a white cane only.

3) Number of collisions involving the white cane: Any contact between any part of the white cane and obstacles or wall dividers is counted as a collision. Our hypothesis was that Virtual Whiskers would reduce the number of white cane collisions compared to using a white cane only.

4) Safety window: The safety window refers to the minimum distance between participants and obstacles. When multiple obstacles exist within a task, the safety window is calculated as the average of the minimum distances between them. Trajectories recorded from the iPhone are overlaid on the obstacle course layout to measure this distance. Our hypothesis was that the safety window would increase for virtual whiskers users compared to using a white cane only.

To minimize counting errors or bias in the experiments, we had three experimenters independently count data points mentioned above. Then, we took the average of their results. Two experimenters, positioned at the starting and ending points, filmed the participants for reference purposes.

We employed a systematic approach to analyzing the performance metrics across different experimental conditions and task difficulties. We employed the Wilcoxon signed-rank test, a non-parametric method, to determine the statistical significance of the differences observed between using Virtual Whiskers and the white cane alone. The significance threshold was set at p¡0.05.

To maintain the integrity of our analysis, we applied a data exclusion criterion to identify and remove potential outliers that could impact the validity of our findings. Outliers were defined as data points falling outside of three standard deviations from the mean for each measurement across the different experimental conditions and difficulty levels. This rule was consistently applied to all measures, ensuring that the analysis remained focused on the central tendencies of the data without being unduly affected by extreme values.

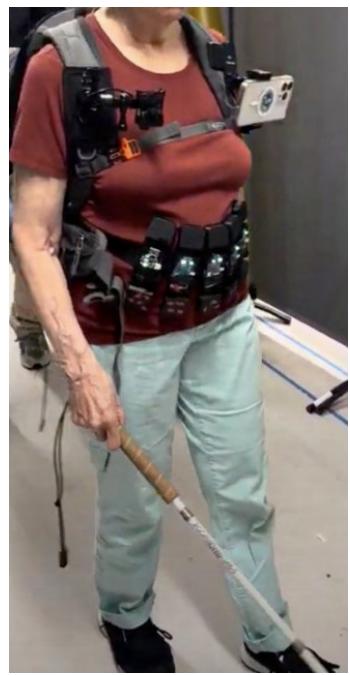

Fig. 10. A person wears the VIS$^4$ION platform and the haptic feedback belt. The haptic feedback belt is worn on the waist. A camera, as a part of the VIS$^4$ION, is mounted on the right shoulder strap of the belt, and an iPhone is mounted on the left shoulder strap of the backpack.

## IV. Results

## V. Participants Demographics

Ten participants with profound visual impairment were recruited for the study, most of whom had a baseline best-corrected visual acuity of no light perception to finger counting. The participants had an average age of 51.3 years. Four of them are identified as female and the rest are identified as male. All participants were familiar with using a long cane. Four of them relied on a white cane as their primary mobility aid, while two used a sighted guide (a friend or family member) and four used a guide dog. All current guide dog users had sufficient experience with the white cane. The two participants who used a sighted guide had less experience with the white cane. The basic demographic information of the participants is presented in Table I.

| Part. ID | Sex | Age | Visual Ability | Prim. Mobility Aid | Cane Use |
|---|---|---|---|---|---|
| p1 | M | 36 | No vision | Guide dog | 30 years |
| p2 | M | 33 | Hand motion | White cane | 26 years |
| p3 | M | 42 | Finger counting | Sighted guide | 0 years |
| p4 | M | 47 | Finger counting | Sighted guide | 0 years |
| p5 | F | 71 | No vision | Guide dog | 35 years |
| p6 | M | 32 | No vision | Guide dog | 27 years |
| p7 | F | 75 | Hand motion | White cane | 15 years |
| p8 | F | 59 | No vision | White cane | 7 years |
| p9 | M | 63 | No vision | White cane | 50 years |
| p10 | F | 55 | No vision | Guide dog | 50 years |

TABLE I
PARTICIPANT DEMOGRAPHICS AND PRIMARY MOBILITY AIDS

### A. Experiment Results

*1) Efficiency measure 1: Completion time:* The first measure is the task completion time.

We visualize the differences in completion time between open path mode vs white only and depth mode vs white cane only, in Figure 11. From the figure, it is noticeable that the majority of the participants experienced an increase in task completion time. For example, 80% of open path users and 90% of depth mode users exhibited an increase in their completion time compared to using the white cane only condition for the easy tasks. For medium and hard tasks, the same pattern was observed as more than half of the participants increased their completion time. The average lines in Figure 11 are more pronounced. Compared to white cane only, open path mode increased average task completion times by 3.4 seconds for easy tasks, 4.7 seconds for medium tasks, and decreased time by 1.8 seconds for hard tasks. In contrast, depth mode increased completion times by 9.9 seconds for easy tasks, 11.9 seconds for medium tasks, and 3.9 seconds for hard tasks. These values indicate that, apart from a slight decrease in the open path scenario for difficult tasks, there is an increase in task completion time across all other cases.

Figure 12 is the box plot of the task completion time for all three groups under three different conditions. We ran Wilcoxon signed-rank tests for the comparisons between Open Path vs. White Cane and depth mode vs. White Cane at each difficulty level.

For the Easy difficulty level, the comparison between depth mode and White Cane shows a significant difference (p = 0.014), indicating that completion times are significantly different between these two conditions. For the Medium difficulty level, the comparison between depth mode and White Cane also shows a significant difference (p = 0.03). Other comparisons (Open Path vs. White Cane for Easy, Medium, and Hard difficulties, and depth mode vs. White Cane for Hard difficulty) do not show significant differences.

*2) Efficiency measure 2: Percent of hesitation time:* We visualize the differences in percent of hesitation time between open path mode vs white only and depth mode vs white cane only, in Figure 13.The percentage of hesitation time reduced in most cases: 70% of open path mode users reduced their percentage of hesitation time in easy tasks, and this number increased to 80% for medium and hard tasks. For depth mode users, the percentage of participants who reduced their hesitation time for easy, medium, and hard tasks was 40%, 50%, and 70%, respectively.

The open path mode significantly reduces the percent of hesitation time in all cases. For depth mode, the average percent of hesitation time was not lower for easy and medium cases but reduced for difficult task. Compared to using a white cane alone, open path mode reduced percent of hesitation time by 7.4%, 8.5%, and 11.3% for easy, medium, and hard tasks, respectively. In contrast, depth mode led to a increase of 1.3% for easy task, and reductions of 0.5%, and 5.2% for the medium and hard task.

Figure 14 is the box plot for the percent of hesitation time. We ran Wilcoxon signed-rank tests for the comparisons between Open Path vs. White Cane and depth mode vs. White Cane at each difficulty level. For the Easy difficulty level, the comparison between Open Path and White Cane shows a significant difference (p = 0.043), indicating that the percent of hesitation time is significantly different between these two conditions. The comparison between depth mode and White Cane does not show a significant difference (p = 0.889). For the Medium difficulty level, the comparison between Open Path and White Cane shows a significant difference (p = 0.038152). The comparison between depth mode and White Cane does not show a significant difference (p = 0.859). For the Hard difficulty level, the comparison between Open Path and White Cane shows a significant difference (p = 0.008). The comparison between depth mode and White Cane shows a marginally significant difference (p = 0.055).

These results suggest that the percent of hesitation time is significantly different between Open Path and White Cane for all difficulty levels, while the comparison between depth mode and White Cane is not significant except for the Hard difficulty level, which shows a marginally significant difference.

*3) Effectiveness measure 1: Number of white cane contacts:* Figure 15 shows the number of white cane contact for each participant. Both modes of Virtual Whiskers reduces the number of cane collisions for most participants. Compared to using a white cane alone, open path mode reduced the average number of white cane contacts by 7.1, 6.5, and 13 times for easy, medium, and hard tasks, respectively. In contrast, depth mode led to reductions of 0.4, 1,8, and 6.4 for the same task difficulties. For open path mode users, the percentage of users with a reduced number of cane collisions is 90%, 100%, and 100% for easy, medium, and hard tasks, respectively. For depth mode users, the percentage of users with a reduced number of cane contacts is 70%, 80%, and 80%, respectively.

Figure 16 shows the box plot for the number of cane contacts. For the easy difficulty level, the calculated Wilcoxon signed-rank tests showed significant differences for the Open Path vs. Cane comparison with a p-value of 0.008, indicating that the open path mode significantly reduces cane collisions compared to the Cane only mode. However, no significant difference was observed for the depth mode vs. Cane comparison (p = 0.176). In the medium difficulty level, significant differences were again observed for the Open Path vs. Cane comparison (p = 0.008), while the depth mode vs. Cane comparison showed no significant difference (p = 0.062). For the hard difficulty level, the Open Path vs. Cane comparison continued to show significant differences (p = 0.008), whereas the depth mode vs. Cane comparison did not show significant differences (p = 0.078).

These findings suggest that the Open Path condition is effective in reducing cane collisions across all difficulty levels, while the depth mode condition does not show significant improvements compared to the Cane only mode. However, the depth mode conditions have marginally significant for both medium difficulty level and hard difficulty level.

Notice that from Figure 15, it is obvious that two participants seem have difficulties in understanding the depth mode. If we exclude participant these two participants' data, for the Easy difficulty level, the comparisons between depth mode and White Cane (p = 0.016) show significant differences. For the Medium difficulty level, the comparisons between the depth mode and White Cane (p = 0.018) show significant differences. For the Hard difficulty level, the comparison between depth mode and White Cane shows a marginally significant difference (p = 0.055). These results suggest that the collision count is significantly different between the depth mode and White Cane for all difficulty levels.

*4) Effectiveness measure 2: Safety Window:* Paths for one representative subject's performance in all nine layouts are shown Figure 17.

For the Easy difficulty level, the comparison between Open Path and Cane does not reveal a significant difference (p = 0.275), indicating that the performance between these two conditions is comparable. Similarly, the comparison between depth mode and Cane does not exhibit a significant difference (p = 0.064), though it is marginally significant, just above the typical threshold.

For the Medium difficulty level, the comparison between Open Path and Cane does not show a significant difference (p = 0.432), suggesting similar performance between these two conditions. The comparison between depth mode and Cane also does not demonstrate a significant difference (p = 0.193).

For the Hard difficulty level, the comparison between Open Path and Cane show a significant difference (p = 0.037), indicating that the performance between these two conditions is significantly different. The comparison between depth mode and Cane, however, does not reveal a significant difference (p = 0.375).

These results suggest that there is a significant difference between Open Path and Cane at the Hard difficulty level, while depth mode shows no significant differences compared to Cane across any difficulty levels.

Figure 18 shows for each task, there always more than half participants have safety window increase under open path mode or depth only mode. Compared to using a white cane alone, open path mode increased the safety window by 5.0cm, 1.4cm, 2.7cm for easy, medium, and hard tasks, respectively. In contrast, depth mode led to increases of 7.3cm, 3.3cm, and 0.75cm for the same task difficulties. Or open path mode, the number of participants with safety window increase is 6, 6, and 7 for easy, medium, and difficult respectively. For depth mode, the number is 7, 6, and 7 for easy, medium, and difficult respectively. It seems that Virtual Whiskers does have a positive effect on increasing the safety window for a majority of participants across all difficulty levels, which is a good indication of its effectiveness. However, whether this effect is statistically significant according to conventional criteria has not been confirmed.

*5) Survey:* We asked all participants about their opinions about Virtual Whiskers after the experiments. If they states they liked both modes of Virtual Whiskers, we coerced them into selecting one mode as the favorite. As the result, four participants preferred depth mode, five participants preferred open path mode, and one participant preferred neither.

## VI. Discussion

Findings from this study highlight the potential of the Virtual Whiskers to enhance mobility and safety of pBLV. Through user testing, we observed that the implementation, particularly the open path mode, resulted in a notable reduction in the number of white cane contacts and hesitation time, across varying difficulty levels. These improvements suggest that Virtual Whiskers effectively aid users in navigating complex environments with fewer obstacles encountered and less uncertainty during travel.

### A. Interpretation of Results

While the average task completion time of all participants increased in either mode across various conditions, this rise was minimal, with no significant increases noted for four out of six tasks. The slight increase in task completion time may be attributed to a heightened cognitive load, as end users leveraged new approaches. Using an ETA requires users to focus on interpreting device feedback alongside navigating, a process that is inherently slower than using the more simple, immediate and intuitive feedback from a white cane. This finding aligns with similar patterns observed in ETA research [24, 35]. Furthermore, considering that participants were first-time users of Virtual Whiskers, it is reasonable to anticipate a reduction in task completion times with repeated use, as users become more accustomed to the system. Future studies should include multiple sessions to monitor how task completion times evolve with increased learning, which could highlight the long-term benefits of the device.

The second efficiency measure evaluated in this study is the percentage of hesitation time. Results demonstrate that the open path mode significantly reduced hesitation time across all tasks, underscoring its potential to enhance navigation efficiency by delivering clear, actionable feedback about the surroundings. This reduction indicates that users were able to make quicker decisions about their path, facilitating smoother navigation. In contrast to relying solely on a white cane, the open path mode offers users information on where to find an alternate route after encountering an obstruction. The depth mode also proved effective, albeit less consistently. Some participants found it more challenging to interpret, which might explain the variability in its effectiveness. However, when excluding the data from the two participants with the poorest performance, the depth mode also shows a significant reduction in hesitation time. This suggests that while the depth mode can be effective, it may require a longer acclimatization period for some users.

Number of cane contacts reflects the effectiveness of virtual whiskers. If Virtual Whiskers was truly effective, it should guide the users around obstacles, prioritizing obstacle-free areas, ergo with less cane contacts. Indeed, we observe that the reduction in cane contacts across all difficulty levels when using Virtual Whiskers, particularly in open path mode, underscoring its potential as an effective tool for obstacle negotiation. This reduction suggests that the device successfully guided users toward safer, more navigable paths, reducing the need for extensive cane exploration. However, the effectiveness of the depth mode was less pronounced, likely due to the complexity of interpreting depth-based feedback. Participants' feedback indicated that the depth mode required more effort to understand, which aligns with the observed data. Excluding data from participants who struggled with the depth mode revealed its potential effectiveness, indicating that with adequate training and familiarization, this mode could also contribute significantly to reducing cane contacts.

The safety window is the second effectiveness measure of virtual whiskers. Although result suggests there is an increase in average safety window in all tasks for both open path and depth mode, compared to white cane only, there is a significant difference between Open Path and Cane at the Hard difficulty level, while depth mode shows no significant differences compared to Cane across any difficulty levels. This might suggest the improvement in safety window is less consistent or perhaps that the easier tasks were too simple and suffered from ceiling effects.

### B. Implications of the Findings

*1) Generalized object detection:* Virtual Whiskers leverages advanced computer vision models, including zero-shot object detection and monocular depth estimation, to provide users with real-time spatial information. A key strength of the system lies in its ability to generalize across different environments without being limited to specific object classes. Unlike traditional ETAs that rely on predefined object classes, Virtual Whiskers' algorithms detect traversable spaces rather than individual obstacles, allowing it to function effectively in a wide range of environments. This approach represents a significant advancement in assistive technology, offering a more versatile and reliable navigation aid for pBLV. However, the effectiveness of this generalized approach can be influenced by environmental factors such as lighting and surface textures, which should be considered in future iterations of the system.

*2) Dual modes operations:* One of the key advantages of the Virtual Whiskers is its dual-mode operation, which offers users the flexibility to select a navigation mode that best suits their personal preferences and situational needs. This higher-order functionality allows for distinct approaches to obstacle negotiation, catering to different user preferences and environmental contexts. This flexibility is particularly beneficial for users with varying levels of experience, comfort, and specific navigation challenges.

During the experiment, we observed a strong preference among participants for one mode over the other. Some participants favored the open path mode for its clear and direct guidance in identifying the most spacious areas, which allowed them to navigate more confidently in environments with fewer obstacles. Conversely, others preferred the depth mode for its ability to provide nuanced feedback about the proximity of obstacles, which they found valuable in more cluttered or unpredictable environments. The ability to switch between modes empowers users to tailor their navigation strategy based on their immediate environment, enhancing both safety and comfort.

This dual-mode capability also highlights the adaptability of Virtual Whiskers to a wide range of real-world scenarios. For instance, users might opt for the open path mode in open or semi-open environments, where identifying a clear route is crucial, while the depth mode might be more suitable in dense, obstacle-rich environments where detailed feedback on obstacle proximity is needed. By accommodating different user preferences and environmental demands, Virtual Whiskers can serve as a more versatile and personalized navigation aid.

Both operational modes of the Virtual Whiskers incorporate a sophisticated set of rules that filter out extraneous information, ensuring that only the most pertinent obstacles are communicated to the user. Rather than bombarding users with details about all surrounding obstacles, the open path mode strategically identifies and communicates the most traversable direction, enhancing navigability in less cluttered environments. Conversely, the depth mode focuses on providing detailed alerts about close, significant obstacles, crucial for navigation in more complex settings. These rules have been meticulously optimized through empirical testing, leading to a system design that reduces unnecessary user disturbance. This selective approach not only minimizes cognitive overload but also improves the overall user experience by enabling more focused and confident navigation.

The observed variation in mode preference underscores the importance of providing customizable options in assistive technologies. It also suggests that future developments could focus on refining the user interface to make mode switching more intuitive, as well as exploring additional modes that could further enhance user experience and safety across diverse settings.

*3) Modularized haptic feedback belt design:* Another significant advantage of the Virtual Whiskers is the modularized design of the haptic feedback. This design addresses several challenges identified in previous prototypes, particularly concerning durability and user convenience. The modular belt design eliminates exposed wires, which were a primary point of failure in earlier versions, especially given the frequent on-and-off usage by users. By encasing all components within durable, self-contained units, the units becomes far less susceptible to breakage, significantly enhancing its longevity. In addition, these smaller units can be donned and doffed on a personal belt, drastically improving comfort and fit.

Moreover, the modular design simplifies repairs, as individual units can be easily replaced without requiring extensive work on the entire wearable. This is a considerable improvement over the wired design, where damage to a single component often necessitated complex repairs or even complete replacement. The ease of maintenance provided by the modular design ensures that the system remains operational with minimal downtime, which is crucial for users who rely on it for daily navigation.

In addition to durability and ease of repair, the modular design offers users the flexibility to customize their belts according to their needs and preferences. Users can choose to add more modular units for greater accuracy and a more immersive experience, or they can opt for fewer units for a more economical setup. This customization capability allows Virtual Whiskers to cater to a broad spectrum of users, from those seeking high-performance navigation aids to those looking for a cost-effective solution. The ability to tailor the system in this way further enhances its appeal and practicality, making it a versatile tool for individuals with varying levels of vision impairment and navigation needs.

*C. Limitations*

The easy, and potentially medium, tasks may have suffered from ceiling effects, during which the simplicity of the task limited the ability to detect differences between Virtual Whiskers and traditional cane use. This effect suggests that the true benefits of Virtual Whiskers might be more apparent in complex, congested environments where navigation challenges are more pronounced. The ceiling effect highlights the potential utility of Virtual Whiskers in crowded or unfamiliar spaces, where its ability to identify and guide users toward safer paths could be particularly valuable. Future studies should focus on environments that more accurately reflect the real- world challenges faced by pBLV, to better assess the practical applications of this technology.

The post-experiment survey revealed a divergence in participant preferences: four participants favored the depth mode, while five preferred the open path mode. Surprisingly, two participants who expressed a preference for depth mode actually demonstrated better performance with the open path mode. This unexpected outcome suggests that individual perceptions of efficacy may not always align with actual performance. Such findings underscore the need for further investigation into how users' preferences relate to their practical experiences and effectiveness with different modes, potentially guiding more tailored and effective implementations of the technology.

*D. Future Research Directions*

One critical area is expanding human subject experiments to assess the system's performance in real-world environments, particularly those with dynamic elements such as moving vehicles and pedestrians. Testing in these more complex and unpredictable settings will provide valuable insights into the system's robustness and adaptability, ensuring that it can reliably support users in a broader range of situations.

Another key focus for future research is understanding the underlying reasons for users' mode preferences. Possibly, references may be influenced by users' past experiences, such as their familiarity with other mobility aids. For instance, users who rely on guide dogs might naturally gravitate toward the open path mode, as it mirrors the way a guide dog directs them into safe, traversable spaces. Similarly, users who are accustomed to using a white cane might find the depth mode appealing, as it closely resembles the way a cane indicates the presence of obstacles. Exploring these connections in more depth could help us refine the system to better align with different users' needs and preferences.

Additionally, there is potential to enhance the Virtual Whiskers by integrating audio prompts alongside the haptic feedback. These prompts could provide users with more explicit directional guidance, such as "turn left in 100 meters", complementing the vibratory signals and further enhancing situational awareness. The combination of haptic and auditory feedback could offer a richer, multi-modal navigation experience, catering to different sensory preferences and improving overall usability.

## VII. CONCLUSION

In this paper, we introduce Virtual Whiskers, a pioneering haptic-based, higher-order sensory substitution system designed to tackle the significant challenges faced pBLV. Our approach integrates advanced computer vision models into two distinct operational modes—open path mode and depth mode—offering users tailored feedback to enhance their spatial awareness and obstacle negotiation capabilities. Extensive experiments were conducted to evaluate the system's performance, demonstrating its effectiveness in reducing hesitation times and the number of white cane contacts, thereby improving navigational safety and efficiency for pBLV.

Virtual Whiskers exemplifies the downstream application of advanced computer vision models into higher-order assistive technology. Practically, our research delivers a robust, customizable system that significantly enhances the mobility and independence of individuals with visual impairments, offering both durability and flexibility through its modularized design. Future research will focus on refining these models' accuracy and expanding user testing to further validate and enhance the system's capabilities.

By advancing the understanding and application of assistive technologies, Virtual Whiskers paves the way for creating more inclusive and accessible environments, ultimately contributing to the broader goal of fostering independence and improving the quality of life for individuals with visual impairments.

ACKNOWLEDGMENT

This research was supported by the National Science Foundation under Grant Nos. ECCS-1928614, CNS-1952180, and ITE-2236097 by the National Eye Institute and Fogarty International Center under Grant No. R21EY033689, as well as by the U.S. Department of Defense under Grant No. VR200130. The content is solely the responsibility of the authors and does not necessarily represent the official views of the National Institutes of Health, National Science Foundation, and Department of Defense.

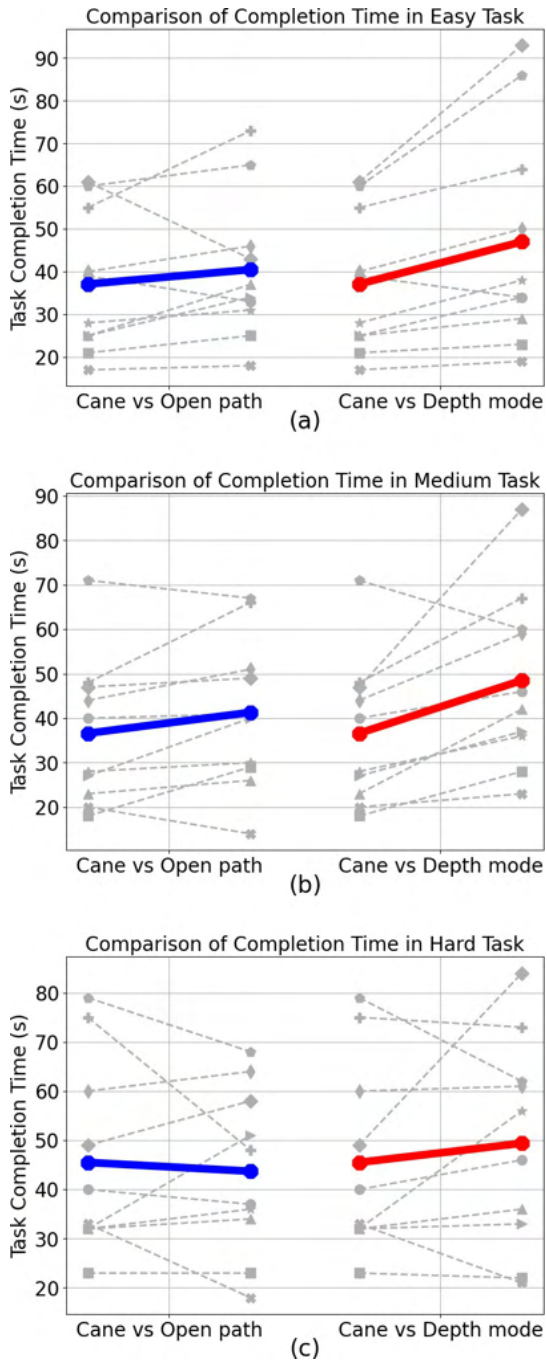

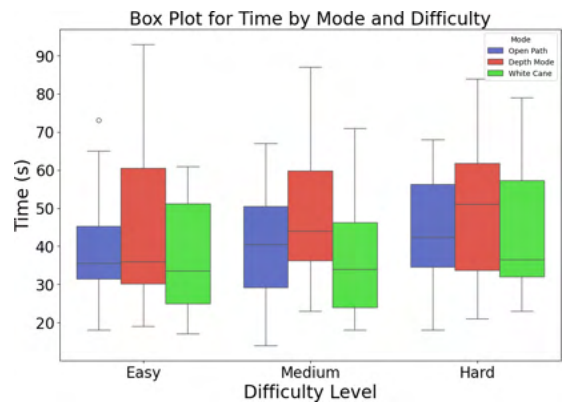

Fig. 12. Box plot for the task completion time. From left to right, the left three boxes are visualizations of task completion time on easy task for open path, depth, and white cane respectively. The middle three boxes are visualizations of task completion time on medium task for open path, depth, and white cane respectively. The right three boxes are visualizations of task completion time on hard task for open path, depth, and white cane respectively. The blue lines and the red lines display the data average of all participants for cane vs open path mode and cane vs depth mode respectively.

Fig. 11. Line plot comparing task completion times across different conditions. On the left side of the plot, each grey line compares the task completion times between using the white cane only and the open path mode. The left end of these lines indicates the completion time under the white cane only condition, while the right end shows the time under the open path condition, for each participant. Similarly, the right side of the plot focuses on the comparison between the white cane only and the depth mode. Here, the left end of each line denotes the completion time for the white cane, and the right end represents the time under the depth mode for each participant. Different markers on the lines identify individual participants. Subplots (a), (b), and (c) represent the completion times for tasks of easy, medium, and hard difficulty levels, respectively. Blue lines across the plot indicate the average data for all participants comparing the white cane and open path mode and red lines represent the averages for the white cane versus depth mode comparisons.

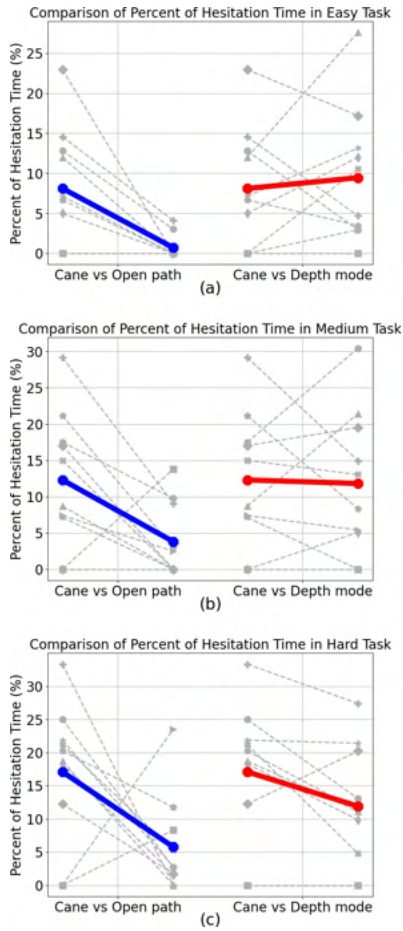

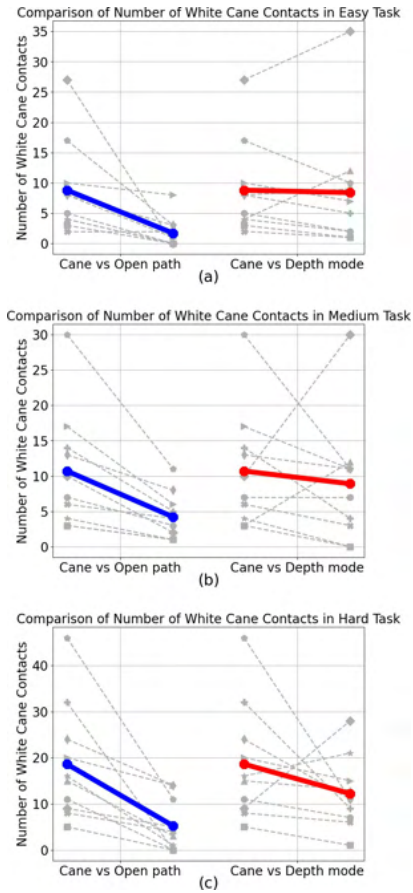

Fig. 13. Line plot comparing percent of hesitation time. Left side: White cane vs. open path mode. Right side: White cane vs. depth mode. Each line's left end shows White cane values, right end shows depth mode values. Different markers on the grey lines signify individual participants. Subplots (a), (b), and (c) represent the percents of hesitation time for tasks of easy, medium, and hard difficulty levels, respectively. Blue lines across the plot indicate the average data for all participants comparing the white cane and open path mode, while red lines represent the averages for the white cane versus depth mode comparisons.

Fig. 15. Line plot comparing number of cane contacts. Left side: White cane vs. open path mode. Each line's left end shows white cane only values, right end shows open path mode values. Right side: White cane vs. depth mode. Different markers on the grey lines signify individual participants. Subplots (a), (b), and (c) represent the number of cane contacts for tasks of easy, medium, and hard difficulty levels, respectively. Blue lines across the plot indicate the average data for all participants comparing the white cane and open path mode, while red lines represent the averages for the white cane versus depth mode comparisons.

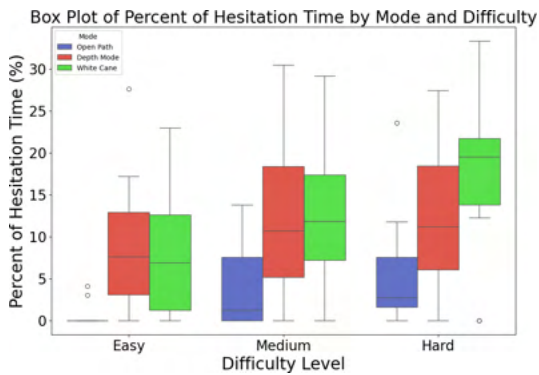

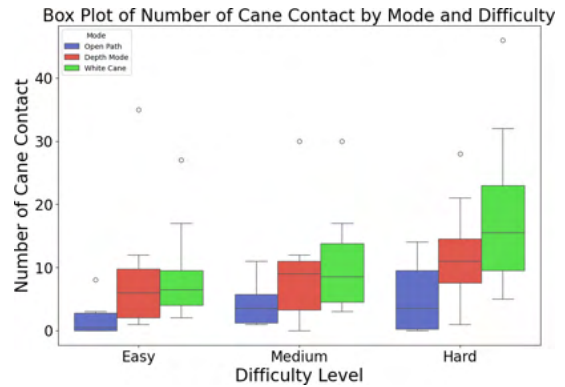

Fig. 14. The box plot for the percentage of hesitation time. Left three boxes: Easy task (Open Path, Depth, White cane). Middle three boxes: Medium task (Open Path, Depth, White cane). Right three boxes: Hard task (Open Path, Depth, White cane)

Fig. 16. The box plot for the number of cane contacts. Left three boxes: Easy task (Open Path, Depth, White cane). Middle three boxes: Medium task (Open Path, Depth, White cane). Right three boxes: Hard task (Open Path, Depth, White cane)

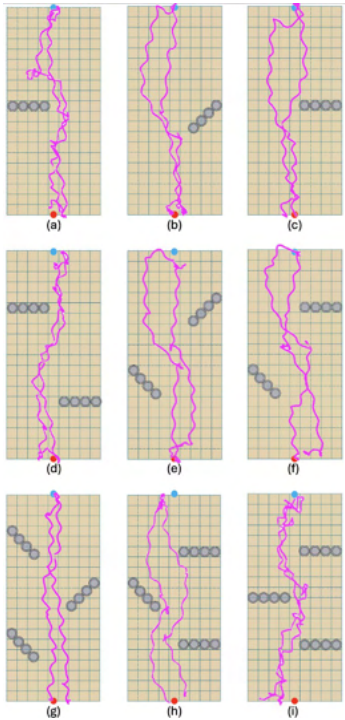

Fig. 17. The subject's walking paths through a hallway using open path mode, depth mode and Only only. (a), (d), and (g) are paths for easy, medium, and hard tasks using open path mode. (b), (e), and (h) are paths for easy, medium, and hard tasks using white cane only. (c), (f), and (i) are paths for easy, medium, and hard tasks using depth mode. This set of trials shows the performance of a representative subject

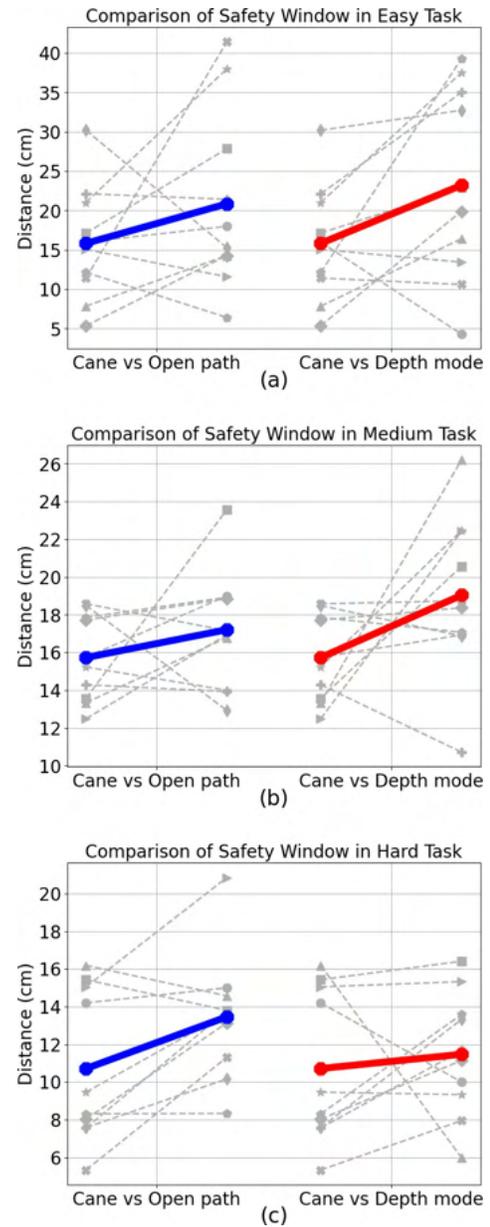

Fig. 18. Line plot comparing safety window. Left side: White cane vs. open path mode. Each line's left end shows white cane only values, right end shows open path mode values. Right side: White cane vs. depth mode. Different markers on the grey lines signify individual participants. Subplots (a), (b), and (c) represent the number of cane contacts for tasks of easy, medium, and hard difficulty levels, respectively. Blue lines across the plot indicate the average data for all participants comparing the white cane and open path mode, while red lines represent the averages for the white cane versus depth mode comparisons